\documentclass{Interspeech}

\interspeechcameraready

\title{MPO: Multidimensional Preference Optimization for Language Model-based Text-to-Speech}

\author[affiliation={1}]{Kangxiang}{Xia}
\author[affiliation={1}]{Xinfa}{Zhu}
\author[affiliation={1}]{Jixun}{Yao}
\author[affiliation={1,*}]{Lei}{Xie}

\affiliation{}{Audio, Speech and Language Processing Group (ASLP@NPU), School of Computer Science, Northwestern Polytechnical University,}{Xi'an, China}
\email{xkx@mail.nwpu.edu.cn, lxie@nwpu.edu.cn\thanks{* Corresponding author.}}


\usepackage{comment}
\usepackage{color}

\begin{document}

\maketitle

\begin{abstract}
    In recent years, text-to-speech (TTS) has seen impressive advancements through large-scale language models, achieving human-level speech quality. Integrating human feedback has proven effective for enhancing robustness in these systems. However, current approaches face challenges in optimizing TTS with preference data across multiple dimensions and often suffer from performance degradation due to overconfidence in rewards. We propose Multidimensional Preference Optimization (MPO) to better align TTS systems with human preferences. MPO introduces a preference set that streamlines the construction of data for multidimensional preference optimization, enabling alignment with multiple dimensions. Additionally, we incorporate regularization during training to address the typical degradation issues in DPO-based approaches. Our experiments demonstrate MPO's effectiveness, showing significant improvements in intelligibility, speaker similarity, and prosody compared to baseline systems\footnote{Speech samples: https://anonymous-person01.github.io/MPO-demo}.
\end{abstract}

\vspace{-6pt}
\section{Introduction}

Recent advancements in text-to-speech (TTS) technology have been impressive, particularly with the development of decoder-only language models (LMs) that generate diverse speech through next-token prediction manner, conditioned on text input. LM-based TTS systems convert speech waveforms into sequences of discrete tokens using neural audio codecs \cite{SoundStream, HFNAC, FunCodec, speechtokenizer, single_codec} and operate in a discrete space \cite{valle, interspeech_2024_chall}. By scaling up both data size and model parameters, LM-based TTS systems have developed emergent in-context learning capabilities, improving their ability to learn the relationships between input text and output speech tokens. 
These systems also demonstrate remarkable zero-shot capabilities in tasks such as voice cloning and cross-lingual synthesis \cite{BASE_tts,VoiceCraft,Seed-TTS}. 

Generating high-quality and natural-sounding speech requires not only scaling up training data \cite{Llasa} but also aligning with human perception \cite{Seed-TTS}. Preference alignment (PA) is a set of training algorithms commonly used in text-based LM development to align model outputs with specific human preferences \cite{Training_lmtfiwhb, Model_aapto}. Typically framed as a reinforcement learning problem, PA first models these preferences using a reward model, and then guides LMs to generate content that maximizes the reward values. When these preferences are derived from humans, the process is called reinforcement learning from human feedback (RLHF) \cite{gpt-4}.

Recent advancements in PA allow for solving the optimization problem in a closed form, eliminating the need for explicit reward modeling \cite{SimPO}, such as Direct Preference Optimization \cite{dpo} (DPO), which significantly simplifies and stabilizes training. Several works in the speech community have explored integrating human evaluation into LM-based TTS optimization. For example, SpeechAlign \cite{SpeechAlign} presents the first method based on DPO that regards ground truth as preferred samples while the generated results as dispreferred samples. UNO \cite{UNO} optimizes unpaired preference data while considering annotation uncertainty in subjective evaluations, RIO \cite{RIO} introduces a reverse preference data selection method based on Bayesian principles. Additionally, some studies have explored screening preference data across multiple evaluation dimensions for preference optimization \cite{Preference_ailmt}. It is also reported that industrial systems, such as SeedTTS \cite{Seed-TTS}, adopt PA in the post-training stage to align the model with human preference.

Despite these advancements, we find two challenges remain. The first is that DPO-based approaches can suffer from performance degradation due to overconfidence in assigning rewards, leading to suboptimal policies \cite{Robust_potrwd}. In extreme cases, this issue will cause the probability of generating the originally preferred token to drop to zero. The second challenge is that directly optimizing TTS systems with preference data across multiple dimensions is difficult. It often requires carefully considering the combined effects of various dimensions when selecting preference data pairs.

In this study, we propose a novel preference optimization approach, called Multidimensional Preference Optimization (MPO), to align TTS systems with human preferences. We introduce a new method for constructing preference datasets, which considers diverse aspects of speech evaluation and enables alignment across multiple dimensions simultaneously. MPO leverages the preference dataset and incorporates additional regularization to address the degradation issues commonly encountered in DPO. Our approach simplifies the construction of preference data and ensures better alignment of synthesized speech with human preferences. Experimental results show that MPO outperforms baseline systems in both subjective and objective evaluations, demonstrating its effectiveness in aligning TTS systems.
The key contributions of this paper are as follows:

\begin{figure*}[h]
\centering
\includegraphics[width=0.8\linewidth]{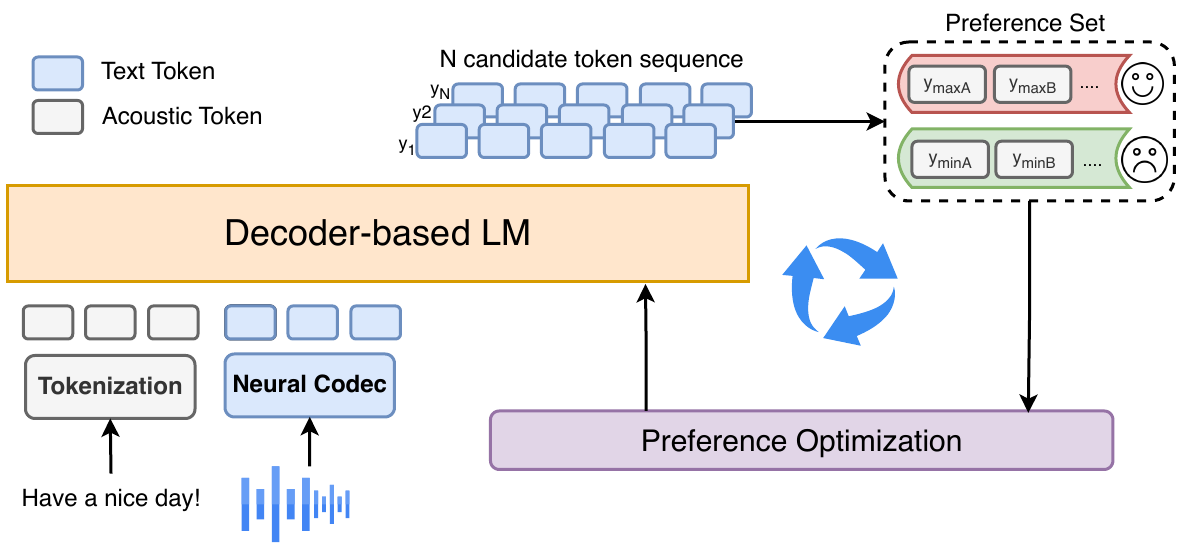}
\vspace{-6pt}
\caption{The overall architecture of the proposed MPO method. 
}
\vspace{-6pt}
\label{fig:methodology}
\vspace{-6pt}
\end{figure*}

\begin{itemize}
    \item We propose a novel preference dataset construction method that captures multiple evaluation dimensions, providing a more comprehensive basis for preference optimization.
    \item We introduce an additional regularization method during preference training to prevent model degradation and ensure stable performance.
    \item We conduct extensive experiments to evaluate the effectiveness of our proposed MPO, showing significant improvements in intelligibility, speaker similarity, and prosody of the generated speech compared to baseline systems.
\end{itemize}

\vspace{-8pt}

\section{Preliminaries}
\vspace{-4pt}
\subsection{Preference Alignment}
\vspace{-6pt}

Preference alignment is often formatted as a reinforcement learning problem. Let \(x\) be the input prompts, and let \(y\) be the language model's response to \(x\).  Given reward function \(r(x, y)\) and reference policy \(\pi_{\text{ref}}\), the goal of alignment is to solve for the "aligned" policy \(\pi_\theta\) that maximizes the excepted reward:
\begingroup
\setlength{\abovedisplayskip}{4pt}
\setlength{\belowdisplayskip}{1pt}
\begin{align}
\label{eq:1}
\max_{\pi_\theta} \mathbb{E}_{y \sim \pi_\theta(y|x)} \left[ r(x, y) \right] &- \beta D_{\text{KL}} \left( \pi_\theta(y|x) \Vert \pi_{\text{ref}}(y|x) \right) 
\end{align}
\endgroup
Here, the KL-divergence term, controlled by the hyper-parameter \(\beta\), prevents the aligned policy from deviating significantly from the reference policy, with a larger \(\beta\) indicating a stronger constraint. However, the reward function \(r\) is usually unknown and is instead constructed from collected human preference data in the form of \((x, y_w, y_l)\), where \(y_w\) is the 'winner', or preferred response, and \(y_l\) is the 'loser', or disfavored response. Given known preference data \((x, y_w, y_l)\), \(r\) can be estimated using the maximum likelihood estimation method:
\begingroup
\setlength{\abovedisplayskip}{4pt}
\setlength{\belowdisplayskip}{1pt}
\begin{align}
\hat{r} \in \arg\min_{r} \mathbb{E}_{(x, y_w, y_l)} \left[ -\log \sigma \left( r(x, y_w) - r(x, y_l) \right) \right]
\end{align}
\endgroup
Here \(\sigma\) is the sigmoid function. With \(\hat{r}\) in hand, policy \(\pi_\theta\) in Eq \ref{eq:1} can be optimized.
\vspace{-6pt}
\subsection{Direct Preference Optimization}
\vspace{-6pt}
Specifically, the optimization in Eq \ref{eq:1} can be solved in closed form without building an explicit reward model. DPO utilizes the form of the optimal solution to the KL-constrained objective to reparameterize the true reward function \cite{dpo}. That is:
\begingroup
\setlength{\abovedisplayskip}{4pt}
\setlength{\belowdisplayskip}{2pt}
\begin{align}
r(x, y) = \beta \log \left( \frac{\pi_\theta(y|x)}{\pi_{\text{ref}}(y|x)} \right) + \beta \log Z(x)
\end{align}
\endgroup
Under the Bradley-Terry\cite{Bradley-Terry} model, the probability that \(y_w\) is preferred over \(y_l\) is given by:
\begingroup
\setlength{\abovedisplayskip}{4pt}
\setlength{\belowdisplayskip}{2pt}
\begin{align}
P(y_w \succ y_l | x) = \sigma \left( \beta \log \left( \frac{\pi_\theta(y_w| x) \pi_{\text{ref}}(y_l | x)}{\pi_\theta(y_l | x) \pi_{\text{ref}}(y_w | x)} \right) \right)
\end{align}
\endgroup
The policy \(\pi_{\hat{\theta}}\) can be directly estimated on the preference data without the need for an intermediate reward model. The objective function \(L_{\text{DPO}}\) can be written as:
\begingroup
\setlength{\abovedisplayskip}{4pt}
\setlength{\belowdisplayskip}{1pt}
\begin{align}
\mathcal{L}_{\text{dpo}} = \mathbb{E}_{(y_w, y_l, x)} \left[ -\log \sigma \left( \beta \log \left( \frac{\pi_\theta(y_w| x) \pi_{\text{ref}}(y_l| x)}{\pi_\theta(y_l| x) \pi_{\text{ref}}(y_w| x)} \right) \right) \right]
\end{align}
\endgroup
where now estimated policy \(\pi_{\hat{\theta}}(y|x)\) is given by \(\pi_{\hat{\theta}}(y|x) \in \arg\min_{\pi_\theta} L_{\text{DPO}}\), maximizing \(P(y_w \succ y_l)\). 

\section{MPO}
Our proposed MPO improves the original DPO approach for TTS tasks by addressing the challenges of multidimensional preference alignment. MPO involves constructing a multidimensional preference dataset and incorporating additional regularization during training to prevent model degradation. The overall architecture of MPO is illustrated in Figure~\ref{fig:methodology}.The preference optimization process begins with the tokenization of input text into discrete tokens. The decoder-based LM then generates the corresponding speech tokens conditioned on the text tokens. 
We adjust the hyperparameters to promote diverse generation, resulting in multiple candidate token sequences. 
Each sequence is evaluated across multiple dimensions to form a preference set, containing both preferred and dispreferred samples. This preference set guides the optimization process, ensuring that the synthesized speech aligns with human preferences.

% In what follows, we detail the key components of our approach: the construction of the multidimensional preference dataset and the regularized DPO training.

\subsection{Multidimensional Preference Set}
DPO trains policies directly on preference data to align results with human preferences. Typically, for a given input, a preference data pair includes both a preferred and a dispreferred response. However, in speech synthesis tasks, there are often many different evaluation dimensions. The outputs generated by the model may have varying strengths and weaknesses across these dimensions. Screening preference data pairs across multiple dimensions requires considering the combined effects of these factors.

To address this, we propose the concept of a preference set, which breaks away from the traditional constraint of having only one preferred and one dispreferred response for the same input. This approach allows for a more flexible and comprehensive consideration of multiple evaluation dimensions.

The construction of the preference dataset is as follows: For a given text input \(x\), assume the output speech generated by the model is \(y_1, y_2, y_3, \ldots, y_n\). Let \(A\) and \(B\) be two evaluation methods, the preference sets can be described as:
\begingroup
\setlength{\abovedisplayskip}{1pt}
\setlength{\belowdisplayskip}{1pt}
\[
w_{\text{set}} = \{ y_{\text{max}A}, y_{\text{max}B} \}
\]
\[
l_{\text{set}} = \{ y_{\text{min}A}, y_{\text{min}B} \}
\]
\endgroup
where \(y_{\text{max}A}\) and \(y_{\text{max}B}\) are the samples of the most preferred according to evaluation methods \(A\) and \(B\), respectively. Similarly, \(y_{\text{min}A}\) and \(y_{\text{min}B}\) are the outputs that are least preferred. During training, we continue to use the data pair approach. We randomly select one data point each from the \(w_{\text{set}}\) and \(l_{\text{set}}\) to form a preference data pair. When there is only one evaluation method \(A\), this setup reduces to the original form of the data pair, where \(w_{\text{set}}\) contains only \(y_{\text{max}A}\) and \(l_{\text{set}}\) contains only \(y_{\text{min}A}\).

During the construction of the preference dataset, there may be cases where the sets of preferred and dispreferred outputs overlap, i.e., \(w_{\text{set}} \cap l_{\text{set}} \neq \varnothing\). 
% In such cases, we need to resolve the conflict by selecting the second-best or second-worst samples for each metrics.
In such cases, we resolve the conflict by selecting the second-best or second-worst samples for one metric. 
Specifically, if an element \(y\) appears in both \(w_{\text{set}}\) and \(l_{\text{set}}\), we replace it in one of the sets by choosing the second most preferred or second least preferred output according to the respective evaluation method. 
% This ensures that the preference dataset remains consistent and representative of the evaluation criteria.
By constructing the preference dataset in this way, we can enhance the contrast between preferred and dispreferred data within preference pairs. This enables the model to optimize more effectively towards human preferences across multiple dimensions simultaneously.
% By constructing the preference dataset in this way, we can effectively capture the nuances of multiple evaluation dimensions and improve the alignment of the model outputs with human preferences.

% \subsection{DPO Training with Additional Regularization}
\subsection{Regularized Training}
\label{section:Regularized}

To address the degradation issues encountered in DPO, MPO incorporates additional regularization during the training phase. DPO relies on the Bradley-Terry assumption, which is sensitive to preference data. If the preference probability for one response over another is 1, it will result in a probability of 0 for the non-preferred response. The global optimal solution of the DPO loss may cause the policy to shift the probability mass to responses not appearing in the training set, or even assign nearly zero probability to the winning responses in the training data. This situation is similar to overfitting and can lead to degradation without additional regularization.

For example, if we have a pair of preference responses \(y_w\) and \(y_l\), the global minimum point of the DPO objective in the form of \(\pi_{\hat{\theta}}\) is achieved if and only if \(P(y_w \succ y_l) = 1\), i.e.
\begingroup
\setlength{\abovedisplayskip}{1pt}
\setlength{\belowdisplayskip}{1pt}
\[
\frac{\pi_{\theta^*}(y_w|x)\pi_{\text{ref}}(y_l|x)}{\pi_{\theta^*}(y_l|x)\pi_{\text{ref}}(y_w|x)} \to \infty
\]
\endgroup
Typically, the reference model \(\pi_{\text{ref}}\) used in DPO is already a model fine-tuned with supervised fine-tuning (SFT). For any \(y\) in the preference dataset, it holds that \(0 < \pi_{\hat{\theta}}(y) < 1\). This means that under these circumstances, any \(\hat{\theta}\) that only satisfies \(\pi_{\hat{\theta}}(y_l) = 0\) and \(\pi_{\hat{\theta}}(y_w) > 0\) for all pairs in the preference dataset is a global minimum point of the DPO objective. Clearly, this issue can be seen as a typical example of overfitting. Unlike overfitting to overly predicted responses in the training set, we might overfit to nearly incomprehensible synthesized audio. Moreover, such degradation will occur easily without additional regularization in typical preference datasets.

Considering that LM-based TTS maximizes the posterior of the target sequence \(y\) through a cross-entropy (CE) objective \(L_{\text{ce}}\), we retained the cross-entropy objective during the DPO training phase to prevent model degradation. Thus, the combined loss function is:
\begingroup
\setlength{\abovedisplayskip}{0pt}
\setlength{\belowdisplayskip}{0pt}
\begin{align}
L = \lambda L_{\text{dpo}} + L_{\text{ce}}
\end{align}
\endgroup
where \(\lambda\) is a hyper-parameter to balance the training process. 
% Experimental results have demonstrated the effectiveness of this approach.

\section{Experimental Results}

\subsection{Dataset}

We train the base language model from scratch using multiple datasets: WenetSpeech4TTS \cite{WenetSpeech4TTS}, LibriHeavy \cite{Libriheavy}, and an internal dataset, totaling 160,000 hours of speech data. The internal dataset is created from web-crawled audio and processed according to the data preparation pipeline described in WenetSpeech4TTS. The base model is then fine-tuned on a 2000-hour TTS dataset, which includes both internal and open-source data \cite{covoc} with more accurate text transcriptions. This fine-tuned model serves as the baseline for our experiments. Preference optimization is conducted on a 100-hour high-quality Mandarin TTS corpus.

\vspace{-6pt}
\subsection{Configuration}
The base language model follows the similar architecture of LLaMA \cite{llama},  predicting acoustic tokens conditioned text input in an autoregressive manner. We employ Byte Pair Encoding (BPE) and a neural codec for text and speech tokenization, respectively. The neural codec model consists of a single quantizer with a codebook size of 8192. The base language model consists of 24 transformer layers with 16 attention heads and the input embedding dimension is set to 1024. 

The base model is trained over 2 million steps using AdamW optimizer with a peak learning rate of \(3 \times 10^{-4}\). During the supervised fine-tuning stage, the model is fine-tuned over 130,000 steps with a peak learning rate of \(5 \times 10^{-5}\). The base model training and fine-tuning processes employ 8 NVIDIA A6000 GPUs, while preference optimization uses a single NVIDIA A6000 GPU. The learning rate of preference optimization is set to \(1 \times 10^{-6}\), and the hyper-parameter \(\lambda\) is set to 10.

% For preference set preparation, we focus on three evaluation metrics: intelligibility (CER), speaker similarity, and prosody (intonation).
\subsection{Preference Set Preparation}
For preference set preparation, we focus on three aspects of human perception: intelligibility, speaker similarity, and prosody.
\begin{itemize}
    \item{\textbf{Intelligibility:}} We use the pre-trained automatic speech recognition model, Paraformer \cite{Paraformer}, as the intelligibility evaluation tool to convert speech into text transcription. The character error rate (CER) is then calculated by comparing these transcriptions to the ground truth transcripts.
    \item{\textbf{Speaker Similarity:}} We utilize WavLM-large fine-tuned on the speaker verification task \cite{seed_sim} to extract representations of the synthesized audio and calculate the cosine similarity with the representation of the real audio \cite{Seed-TTS}.
    \item{\textbf{Prosody:}} We use Log F0 root mean square error (RMSE) \cite{ESPnet2} to calculate the difference in the log F0 sequences between the generated and reference speech. Dynamic time warping is employed to align the generated and reference speech features of different sequential lengths, following the evaluation script in ESPnet \cite{ESPnet2}.
\end{itemize}
Using the transcripts from the 100-hour high-quality TTS dataset, we generate 10 batches of speech data with the supervised fine-tuned model. We then construct the preference set based on CER, speaker similarity, and prosody metrics. Rather than simply selecting the best and worst results across the three dimensions for each text input corresponding to the 10 synthesized audio samples, we apply specific constraints: the preferred audio must have a CER of 0; the score difference in speaker similarity between the preferred and dispreferred audio must be at least 0.1; and the score difference in prosody must also be at least 0.1. These constraints ensure that the preference dataset reflects significant differences in the evaluation metrics, providing a robust basis for optimizing the model.

\subsection{Effect of Additional Regularization}

For ease of comparison, we only use the preference data filtered by CER from the preference set for the experiments in this subsection. We compared the overall loss changes of the model with and without the CE loss constraint during optimization.
% In Section 3.2, we proposed introducing the CE loss to prevent model degradation during the DPO training phase. 
% We compared the overall loss changes of the model with and without the CE loss constraint during optimization.
\begin{figure}[h]
    \vspace{-6pt}
    \centering

    \includegraphics[width=\linewidth]{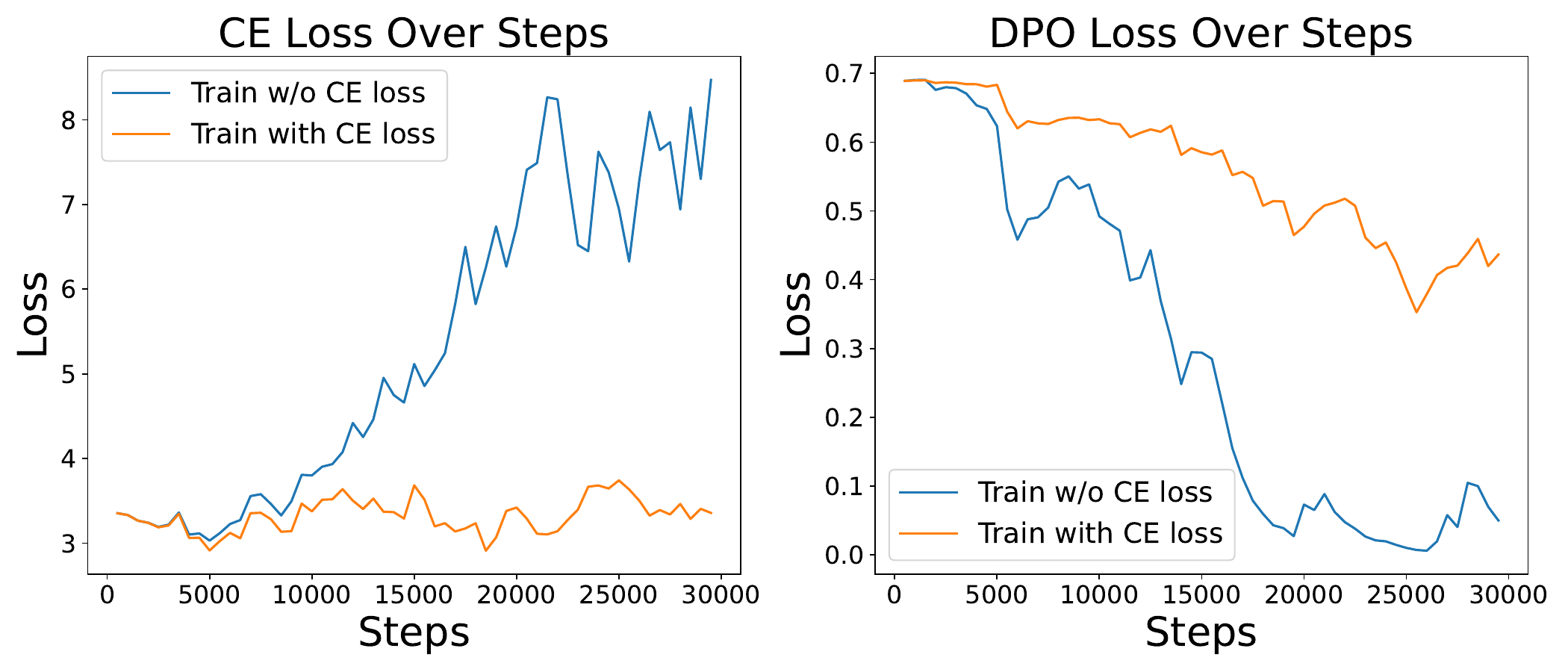}
    \vspace{-6pt}
    \caption{Comparison results of training loss over different training steps.}
    \label{fig:comparison}
    \vspace{-12pt}
\end{figure}

As shown in Figure \ref{fig:comparison}, in the later stages of training, the DPO loss of the model trained without the CE loss constraint nearly converges to zero. At this point, the CE loss also rises to around 10, which is almost the same as the loss in the initial pretraining state, indicating that the model has lost its speech synthesis capability. This outcome aligns with the expected degradation results described in Section \ref{section:Regularized}. Conversely, the model trained with the CE loss constraint does not exhibit signs of degradation.

\begin{table}[bth]
\centering
\vspace{-6pt}
\caption{CER results over different training stages.}
\label{tab:cer_results}
\vspace{-6pt}
\resizebox{0.95\linewidth}{!}{
\begin{tabular}{lccccc}
\toprule
\multicolumn{1}{l|}{\textbf{Model}} & \multicolumn{1}{l|}{SFT model} & 5k steps                 & 10k steps                 & \multicolumn{1}{l|}{15k steps} & Ours \\ 
\midrule
\multicolumn{1}{l|}{\textbf{CER}}   & \multicolumn{1}{c|}{4.72}      & \multicolumn{1}{c}{4.57} & \multicolumn{1}{c}{6.41} & \multicolumn{1}{c|}{14.52}     & 4.24 \\ 
\midrule
\end{tabular}
}
\vspace{-12pt}
\end{table}
To further quantify the effect of the CE loss constraint, we compared the CER results of the models on the test set at different training stages, as shown in Table \ref{tab:cer_results}. The table provides the CER of the supervised SFT model, which is the starting point for subsequent training. It also includes models trained without the CE loss constraint for 5k, 10k, and 15k steps, and our model trained with the CE loss constraint for nearly 20k steps.
From the table, we observe that the CER of the SFT model is 4.72. After 5k steps, the model trained without the CE loss constraint shows a slight improvement with a CER of 4.57. However, as training progresses to 10k and 15k steps, the CER significantly worsens to 6.41 and 14.52, respectively, indicating severe model degradation. In contrast, our model trained with the CE loss constraint achieves a CER of 4.24, demonstrating its effectiveness in preventing degradation.

\vspace{-6pt}
\subsection{Effect of Preference Set}

Under the CE loss constraint, we continue to conduct separate experiments for speaker similarity and prosody. Referring to previous work \cite{Preference_ailmt}, we combine the ranking results of the three evaluation metrics in a naive way. For each metric, we rank all examples and assign scores from 0 to 9, where lower scores indicate better performance. Examples with lower overall scores are preferred.
We compare the models trained using this ranking method with the models trained on the preference dataset we proposed.
\begingroup
\begin{table}[th]
  \caption{Objective evaluation results between baseline systems and our proposed MPO.}
  \vspace{-6pt}
  \label{tab:example2}
  \centering
  \resizebox{0.98\linewidth}{!}{
  \begin{tabular}{lccc}
    \toprule
    \textbf{Model} & \textbf{CER↓} & \textbf{SPK\_SIM↑} & \textbf{Prosody↓} \\
    \midrule
    Ground truth & 7.246 & - & - \\
    \midrule
    Base model & 4.72 & 0.548 & 0.337 \\
    Train on CER & 4.24 & 0.549 & 0.322 \\
    Train on SIM & 5.50 & 0.576 & 0.283 \\
    Train on Prosody & 4.86 & 0.537 & 0.237 \\
    Train on combing rankings & 4.30 & 0.564 & \textbf{0.218} \\
    MPO & \textbf{3.90} & \textbf{0.577} & 0.279 \\
    \bottomrule
  \end{tabular}
  }
  \vspace{-10pt}
\end{table}
\endgroup

As shown in Table~\ref{tab:example2}, applying DPO using any single metric for preference selection results in noticeable improvement primarily in that specific metric. For the two methods that use combined evaluation metrics, the model trained using combined rankings shows relatively average optimization results. In contrast, the model trained with the preference set outperforms the former in both CER and speaker similarity. This is because the preference set more effectively highlights the differences between preferred and dispreferred responses across evaluation dimensions.

\begin{figure}[bth]
  \vspace{-10pt}
  \centering
  \includegraphics[width=\linewidth]{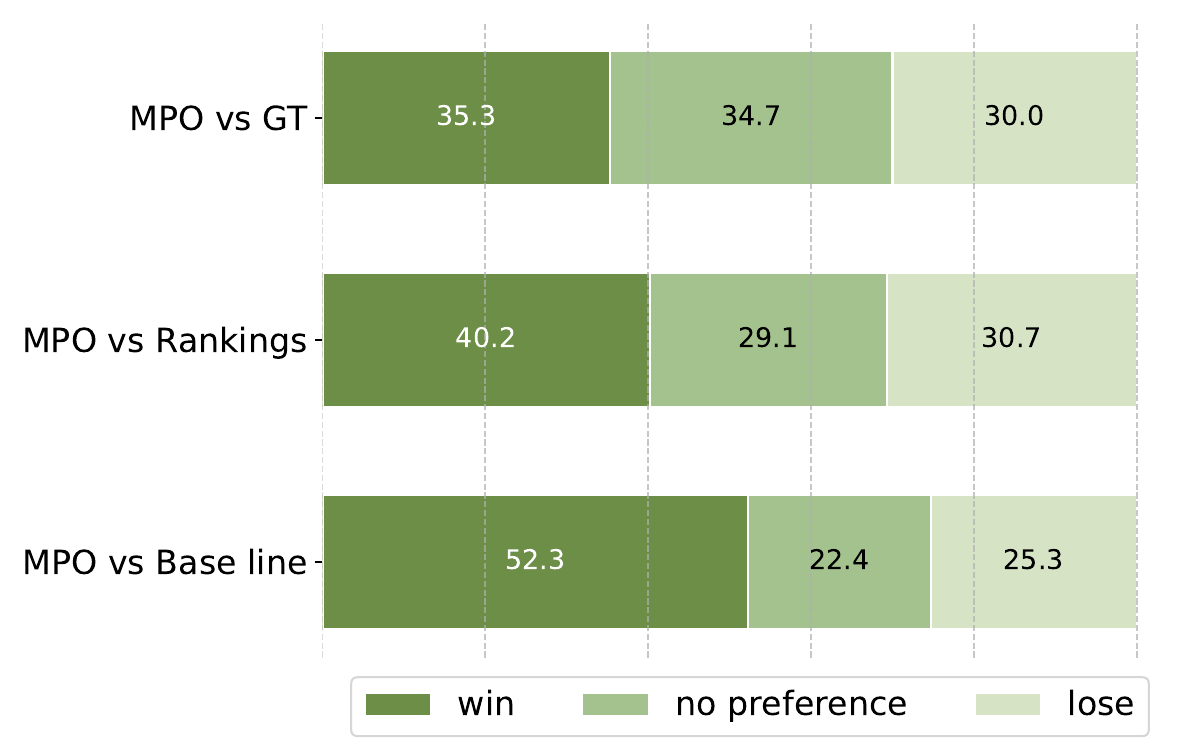}
  \vspace{-10pt}
  \caption{Results of ABX preference test.}
  \label{fig:abx_preference}
  \vspace{-12pt}
\end{figure}

To further verify the overall effectiveness of our proposed training method, we conducted a subjective ABX preference test. The results, illustrated in Figure \ref{fig:abx_preference}, demonstrate the advantages of using the preference set. The figure compares the performance of our MPO method against models trained using combined rankings, GT, and the baseline model.

As shown in Figure \ref{fig:abx_preference}, compared to the baseline model, MPO is preferred in 52.3\(\%\) of cases and ties in 22.4\(\%\), significantly outperforming the base model in preference. Notably, MPO also surpasses models that simply use combined rankings as the basis for optimization (40.2\(\%\) vs. 30.7\(\%\), with 29.1\(\%\) ties), demonstrating the effectiveness of MPO on multidimensional optimization.Additionally, MPO achieves scores comparable to the ground truth, indicating that aligning preferences across the three dimensions results in outputs that better match human preferences.

\vspace{-10pt}
\section{Conclusion}
In this study, we proposed a novel approach, MPO, to enhance the alignment of TTS systems with human preferences. Our method introduces the concept of a preference set, which facilitates the construction of data for multidimensional direct preference optimization, allowing TTS systems to consider multiple evaluation dimensions simultaneously. Additionally, we incorporate regularization during training to address the typical degradation issues observed in DPO-based approaches.
Our experimental results demonstrate significant improvements in intelligibility, speaker similarity, and prosody of the generated speech compared to baseline systems. Specifically, the MPO method outperforms traditional single-metric optimization approaches and combined ranking methods, achieving better alignment with human preferences and producing output that is comparable to ground truth in subjective evaluations.

\bibliographystyle{IEEEtran}
\bibliography{mybib}

\end{document}